\RequirePackage{fix-cm}

\documentclass[twocolumn,epjc3]{svjour3}

\usepackage[english]{babel}
\usepackage{microtype}
\smartqed  

\RequirePackage{graphicx}
\RequirePackage{mathptmx}      
\RequirePackage{flushend}

\RequirePackage[numbers,sort&compress]{natbib}
\RequirePackage[colorlinks,citecolor=blue,urlcolor=blue,linkcolor=blue]{hyperref}
\usepackage{amsmath}

\usepackage{amsfonts}
\usepackage{amssymb}
\usepackage{epstopdf}
\usepackage{slashed}
\usepackage{dsfont}
\journalname{Eur. Phys. J. C}
\def\II{\hbox{{1}\kern-.25em\hbox{l}}}

\begin{document}

\title{Correction exponents in the chiral Heisenberg model at $1/N^2$:\\
singular contributions and operator mixing.
}

\author{Alexander N. Manashov\thanksref{e1,addr1,addr2}
        and
        Leonid A. Shumilov  \thanksref{e2,addr1} 
}

\thankstext{e1}{e-mail: alexander.manashov@desy.de}
\thankstext{e2}{e-mail: leonid.shumilov@desy.de}

\institute{II. Institut f\"ur Theoretische Physik, Universit\"at Hamburg, Luruper Chaussee 149, D-22761 Hamburg,
Germany\label{addr1}
          \and
          Institut f\"ur Theoretische Physik, Universit\"at  Regensburg, D-93040 Regensburg, Germany\label{addr2}
}

\date{}

\maketitle

\begin{abstract}
We calculate the  correction exponents in the chiral Heisenberg model in the $1/N$ expansion. These exponents are related to the slopes
of $\beta$ functions at the phase transition point. We present the results at order $1/N^2$  and check that they agree with the results
of the $\epsilon$ expansion near $d = 4$. We find that one of the correction exponents diverges as $d \to 3$. We argue that the
appearance of the pole is a rather general phenomenon and is associated with operator mixing involving  the system of four-fermion
operators. After analyzing the operator mixing structure, we propose a resummation procedure which modifies the exponents already at
\emph{leading order}. We also perform calculations directly in the three-dimensional model and find complete agreement with the resummed
exponents.
\end{abstract}

\allowdisplaybreaks
\section{Introduction}
According to the universality hypothesis diverse physical systems that share essentially the same symmetry properties exhibit
identical physical behavior close to their critical points~\cite{Kadanoff:1966wm}.
The best-known example of this type is the critical equivalence of  the three-dimensional Heisenberg ferromagnet and the scalar $\varphi^4$
field theory~\cite{Wilson:1973jj}.

In recent years, a lot of attention has been paid to the so-called Gross-Neveu-Yukawa universality
class~\cite{Mihaila:2017ble,Zerf:2017zqi,Gracey:2017fzu,Gracey:2018qba,Gracey:2025aoj}.
 The best-known representative of this class is the Gross-Neveu model \cite{Gross:1974jv} which describes
a system of fermion fields with quartic interaction. It is renormalizable in two dimensions, asymptotically free and, moreover, admits an
exact solution~\cite{Zamolodchikov:1978xm}. In $2<d<4$ dimensions the $N$ component GN-model can be analyzed  using large $N$ expansion
technique. The UV completion  of the GN model contains an additional scalar field  with a quartic self-interaction and is known as the
Gross-Neveu-Yukawa (GNY) model \cite{ZinnJustin:1991ksq}. The critical indices in these models are known with high accuracy in
$2+2\epsilon$, $4-2\epsilon$ and  $1/N$
expansions~\cite{Gracey:1990sx,Gracey:1991vy,Karkkainen:1993ef,Mihaila:2017ble,Zerf:2017zqi,Ladovrechis:2022aof,Gracey:2025aoj,Huang:2025ree,%
Gracey:1990wi,Gracey:1992cp,Vasiliev:1992wr,Gracey:1993kc,Vasiliev:1993pi,Gracey:2017fzu,Manashov:2017rrx}.

The chiral Heisenberg model (CH) is a natural  extension of the GNY model. It describes a system of   scalar fields, $\pi^a$, $a=1,2,3$,
interacting with a  doublet of $N$-component fermion fields, $q^{i,\alpha}$, $\alpha=1,2$, $i=1\ldots N$. This model  is believed to
underlie critical phenomena in graphene, see e.g.
refs.~\cite{Khveshchenko_2001,Herbut:2006cs,Honerkamp_2008,Herbut:2009qb,Raghu_2008,Maciejko:2013lua,Kobayashi_2014,%
Louvet:2016kwp,Syzranov:2015lnb}. The basic critical indices in the CH model were calculated  with  four-loop accuracy  in $4-2\epsilon$
expansion~\cite{Zerf:2017zqi}.  In the $1/N$ expansion the basic indices  are known at order $1/N^2$, and the index $\eta$ -- the anomalous
dimension of the fermion field -- at  $1/N^3$, see ref.~\cite{Gracey:2018qba}.  The results of $1/N$ calculations and calculations in
$\epsilon$ expansion near $d = 4$ are in complete agreement with each other.

In this paper, we calculate  two  other indices -- the so-called correction exponents. They are related to the slopes of $\beta$ functions
at the critical point and so far have been known only at order $1/N$~\cite{Gracey:2018qba}. In our analysis, we heavily use the results of
refs.~\cite{Vasiliev:1993pi,Manashov:2017rrx} and obtain the correction exponents in the CH model at order $1/N^2$. In addition, we
calculate the anomalous dimensions of the operators $\pi^m$ for arbitrary $m$ and $\pi^{\{a}\pi^{b\}} = \pi^a\pi^b -
\tfrac{1}{3}\delta^{ab}\pi^2$ at order $1/N^2$. When expanded in $\epsilon$, our results are in complete agreement with the results
of~\cite{Zerf:2017zqi}. However, one of the correction exponents at order $1/N^2$ diverges when  $d \to 3$. We argue that this is a
rather general phenomenon caused by the different structure of operator
 mixing between $d\neq 3$ and $d=3$.  The presence of a pole at order $1/N^2$ results in a modification
of the leading  order indices in $d=3$.

The paper is organized as follows: In sect.~\ref{sect:CH}, we recall the formulation of the CH model and discuss  the relation of the
slopes of $\beta$ functions  with the critical dimensions of  certain operators. In  sect.~\ref{sect:largeN}, we review briefly the $1/N$
expansion techniques. Section~\ref{sect:omegas} contains some details of the calculations as well as the final results for the correction
exponents in $d$ dimensions. In section~\ref{sect:dto3}, we discuss the behavior of the  critical dimensions in the limit $d \to 3$.
Section~\ref{sect:summary} contains the concluding remarks. The appendices are devoted to technical details of the calculations. For the
readers' convenience all critical indices, including the correction exponents at order $1/N^2$, are collected in the ancillary file.

\section{The chiral Heisenberg model in $d=4-2\epsilon$ dimensions}\label{sect:CH}
The Lagrangian of the CH model in $d=4-2\epsilon$ Euclidean space takes the form
\begin{align}\label{CHd4}
\mathcal{L}=
\frac12(\partial\pi)^2+\bar q\slashed{\partial} q+g_1  \bar q\widehat \pi q+ g_2(\pi^2)^2,
\end{align}
where $q^{i,\alpha}$ is an $N$-component, $i=1,\ldots,N$, $\alpha=1,2$, fermion field and $\pi^a$ is a scalar triplet. Here and below
\begin{align*}
\widehat\pi=\sum_a \sigma^a \pi^a,&& \pi^2=\sum_a(\pi^a)^2, && \pi^4\equiv(\pi^2)^2
\end{align*}
etc. Here $\sigma^a$, $a=1,2,3$,  are Pauli matrices.

 The model is multiplicatively renormalizable~\cite{Zerf:2017zqi}
\begin{equation}\label{eq:ren_action}
 \mathcal{L}_R =\frac{Z_1}2(\partial\pi)^2 +Z_2\bar q\slashed{\partial}q
+M^\epsilon Z_3 g_1 \bar q\widehat\pi  q
+M^{2\epsilon}Z_4 g_2\pi^4.
\end{equation}
Here $M$ is the renormalization scale and $Z_i$ are the renormalization factors which, in the MS-like scheme, are series in $1/\epsilon$,
$Z_i=1+\sum_k z_{i}^k/\epsilon^k$.

The model possesses nontrivial fixed points,
\begin{align}
  \beta_1(g_1^\ast,g_2^\ast) = 0 , &&
  \beta_2(g_1^\ast,g_2^\ast) = 0,
\end{align}
where, as usual
$\beta_k(g_1,g_2)= M\dfrac{dg_k}{dM}\,.$
One of these points is infrared stable, i.e. the eigenvalues $\omega_a$ of the matrix
\begin{align}
\omega_{ik}=\partial_i {\beta}_k\big|_{g=g^\ast}
\end{align}
are both positive. These eigenvalues are called correction exponents.

These indices can be identified as the anomalous dimension of certain composite operators.
The standard RG analysis, see e.g.~\cite{Vasilev:2004yr,Manashov:2017rrx} shows that the correction exponents $\omega_a$ are
related to the critical dimensions of the operators
\begin{align}
\mathcal{O}_A =\bar q \pi q,  &&  \mathcal{O}_B = \pi^4 .
\end{align}
The operators $\mathcal{O}_k$ mix under renormalization and their scaling dimensions are determined by the matrix,
\begin{align}
\Delta & =\Delta_\text{can}+\gamma_\ast
= d+\begin{pmatrix}
    \partial_{g_1} \beta_{1} & M^{\epsilon} \partial_{g_1} \beta_{2} \\
     M^{-\epsilon} \partial_{g_2} \beta_{1} &  \partial_{g_2} \beta_2
  \end{pmatrix}\,.
\end{align}
Diagonalizing the matrix $\Delta$ one constructs two operators with the scaling dimensions
\begin{align}
\Delta_a =d +\omega_a=4+O(1/N), && a=\pm\,.
\end{align}
Being physical observables, scaling dimensions of  operators do not depend on regularization, expansion, etc. We use this property and
calculate the critical indices $\omega_a$ in the $1/N$ expansion.

\section{Large $N$ expansion for the CH model}\label{sect:largeN}
It can be shown that in $d\equiv 2\mu<4$ the CH model~\eqref{CHd4}  is critically equivalent  to the following model~\cite{Gracey:2018qba}
\begin{align}\label{N-CH}
S_\text{CH} &=\int d^d x\left[\bar q\slashed{\partial} q+  \bar q\widehat \pi q -\frac{N}{2g} \pi^2\right]\,,
\end{align}
i.e. the critical indices in both models coincide. In the model~\eqref{N-CH}, critical exponents can be calculated as series in $1/N$.
Below we briefly outline the  technique used for  the calculations. A more detailed overview of the $1/N$ methods, including the
self-consistency equations and the conformal bootstrap, can be found in~\cite{Vasilev:2004yr}.

In the model~\eqref{N-CH} the dominant contribution to the propagator of the $\pi$-field
in the infrared region (IR) comes from the fermion loop~\cite{ZinnJustin:1991ksq} resulting in
\begin{align}\label{sigma-prop}
D^{ab}_\sigma(x)=-\frac{\delta^{ab}}{2n} {B(\mu)}/{x^2}, && n=N\,\text{tr} \II,
\end{align}
where the trace is taken in the space of $d$-dimensional spinors. The amplitude $B(\mu)$ reads
\begin{align}\label{bmu}
B(\mu)=\frac{4\Gamma(2\mu-1)}{\Gamma^2(\mu)\Gamma(\mu-1)\Gamma(1-\mu)}.
\end{align}
The model~\eqref{N-CH} is critically equivalent to a simpler model
\begin{align}\label{L-GN}
S_\Delta &=\int d^d x\left[\bar q\slashed{\partial} q -\frac12 \pi L_\Delta\pi + \bar q \widehat\pi q  + \frac12 \pi L\pi \right]
\end{align}
introduced in~\cite{Vasiliev:1975mq}.
The kernel $L$  is chosen to cancel the fermion-loop contribution to the propagator of $\pi$ field,
\begin{flalign}
L^{ab}(x) & = \mathrm{tr}\, \Big(\sigma^a D_q(x)\sigma^b D_{q}(-x)\Big)
=-\delta^{ab}
\frac{2n A^2(\mu)}{x^{2(d-1)}}\,.
\end{flalign}
Here $D_q$ is the fermion propagator
\begin{align}
D^{\alpha\beta}_q(x) =-\delta^{\alpha\beta}
\frac{A(\mu)\slashed{x}}{x^{2\mu}} \,, &&A(\mu)=\frac{\Gamma(\mu)}{2\pi^\mu}\,.
\end{align}
The regularized kernel $L_\Delta$ is (we suppress isotopic indices)
\begin{align}\label{Ldef}
L_\Delta(x)= L(x) (M^2 x^2)^{-\Delta} C(\Delta)\sim x^{-2(d-1+\Delta)}\,.
\end{align}

The first two terms in~\eqref{L-GN} give  the free part of  the action, $S_0$, and the rest is $S_\text{int}$.
The leading order  propagator of the $\pi$ field
is given by  $L_\Delta^{-1}$. We fix the constant $C(\Delta)$ by a requirement
for the propagator $D_\sigma$ to have the  form
\begin{align}
D^{ab}_\sigma(x)=-  \frac{\delta^{ab}}{2n} B(\mu)(M^{2}x^2)^\Delta/{x^2}\,.
\end{align}
The divergences in the correlators appear as
poles in the  parameter $\Delta$ and are removed by adding  counterterms to the action~\eqref{L-GN}. The renormalized action takes the form
\begin{equation}\label{L-GN-ren}
S_{\Delta, R} =\int d^d x\left[Z_1\bar q\slashed{\partial} q -\frac12 \pi L_\Delta\pi + Z_2  \bar q\widehat\pi q
+ \frac12 \pi L\pi\right].
\end{equation}
The model is renormalizable, however the renormalization is not multiplicative~\cite{Vasiliev:1975mq,Vasilev:2004yr},
 i.e.  $S_{\Delta, R}(q,\pi) \neq S_{\Delta}(q_0,\pi_0)=S_{\Delta}(Z_qq,Z_\pi \pi)$.
As a consequence,  the anomalous dimensions of the fields and operators cannot in general be extracted from  the corresponding
renormalization factors. Fortunately, this effect starts only from  the order $1/n^3$.
Up to the order $1/n^2$ the anomalous dimension of the operators (fields) can be obtained  as
follows~\cite{Vasiliev:1993ux,Derkachov:1997ch}:
\begin{itemize}
\item  replace $D_\pi \mapsto u D_\pi$, where $u$ is an auxiliary parameter. Using the modified propagator,  calculate the
    renormalization factors for the system of operators $\mathcal{O}_i$ mixing under renormalization.
\begin{equation}
[\mathcal{O}_i(\varphi)] =(\mathbf{Z}(u))_{ik} \mathcal{O}^B_k(\varphi_0)\,,
\ \ \ \
 \mathbf{Z}(u) =\sum_{a=0}^\infty \frac{Z_a(u)}{\Delta^{a}},
\end{equation}
where $\varphi=\{q,\bar q, \pi\}$, $Z_0=1$ and $\mathcal{O}^B_i$ are bare operators.
\item The anomalous
dimension matrix for the operators in question can be obtained as follows~\cite{Derkachov:1997ch}
\begin{align}\label{gamma-O}
\gamma_\varphi &=-2u\partial_u \mathbf{Z}_{\varphi, 1}(u)\Big|_{u=1} +O(1/n^3)\,,
\notag\\
\gamma_{ik} &= 2u\partial_u (\mathbf{Z}_1(u))_{ik}\Big|_{u=1} +O(1/n^3)\,.
\end{align}
\end{itemize}
We use these formulae to calculate the correction exponents at order $1/n^2$.

The basic index $\eta=2\gamma_q$ is known with an accuracy of $1/n^3$,  and the anomalous dimensions
\begin{align}
\gamma_\pi=-\eta-2\chi &&\text{ and } &&
\gamma_{\pi^2}\equiv -2\lambda,
\end{align}
with $1/n^2$
accuracy~\cite{Gracey:2018qba}~\footnote{The expressions
for $\eta_{2}$ and $\gamma_{\pi,2}$
in ref.~\cite{Gracey:2018qba} contain typos, however, the expressions in the ancillary file of~\cite{Gracey:2018qba} are correct.}.
Here we present only the leading order expressions,
$\eta=\eta_1/n+\eta_2/n^2+\ldots$, etc., where $n$ is defined in Eq.~\eqref{sigma-prop}
\begin{align}
\eta_1 &= \frac{6}\mu\frac{a(2-\mu)}{a^3(1)}
=\frac{6\Gamma(2\mu-2)}{\mu\Gamma^3(\mu-1)\Gamma(2-\mu)}\,,
\end{align}
where
$a(x)\equiv \Gamma(\mu-x)/\Gamma(x) $
and
\begin{align}
\chi_1&=-\frac{\eta_1}6\frac{\mu}{\mu-1}\,, && \gamma_{\pi, 1}=-\frac{\eta_1}{3}\frac{2\mu-3}{\mu-1}\,,
\notag \\[1mm]
\lambda_1&=-{\eta_1}(2\mu-1)\,.
\end{align}

\section{Correction exponents}\label{sect:omegas}
As was shown in sect.~\ref{sect:CH}  the correction exponents $\omega_a$  in the  $4-2\epsilon$ expansion are related to  the scaling
dimensions of the operators $\pi^4$ and $ \bar q \pi q$, $\Delta_a =4+O(1/n)$. Thus, in the $1/n$ expansion we have to look for the scalar
operators of canonical dimension $4$ (for arbitrary $d$). Taking into account that the canonical dimensions of the fields are
\begin{align}
\dim q=(d-1)/2 &&\text{and} && \dim \pi =1
\end{align}
 these operators
are easy to identify. There are three scalar operators of dimension four,
\begin{align}
\mathcal{O}=\{\pi\partial^2\pi,\pi^4,  \partial^2 \pi^2\}.
\end{align}
The last one is a total derivative and can be neglected. Its anomalous dimension is $\gamma_{\pi^2}$. Therefore, it is enough to consider
only first two operators
\begin{align}
\mathcal{O}_{-}={\pi\partial^2\pi} &&\text{and }&& \mathcal{O}_{+}=\pi^4.
\end{align}
These operators mix under renormalization. The diagonal entries of  the anomalous
dimension matrix and  $\gamma_{12}$ are of the order $1/n$ while  the matrix element $\gamma_{21}\sim 1/n^3$~\cite{Manashov:2017rrx}. Thus
the mixing  does not affect the anomalous dimensions at order $1/n^2$ and can be neglected.

\subsection{$1/n$ results}\label{sect:1n}
%
\begin{figure}[t]
  \centering
  \includegraphics[width=0.82\columnwidth]{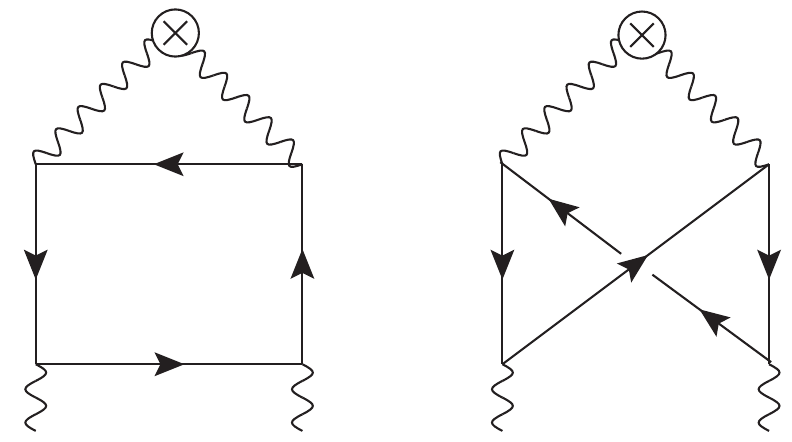}
  \caption{Leading order diagrams.
}
  \label{fig:LO}
\end{figure}

Together with the operators $\mathcal{O}_\pm$  we consider  a more general set of the operators: the scalar operators
\begin{align}
\label{op-number}
\mathcal {O}_m= (\pi^2)^m, && (\mathcal{O}_+
=
 \mathcal{O}_{2})
 \end{align}
and the traceless symmetric operators
\begin{align}
 \widetilde{\mathcal{O}}_{\ell}
 =\pi^{a_1}\ldots
\pi^{a_\ell} -\text{traces}.
\end{align}
The anomalous dimensions of the operators $\widetilde{\mathcal{O}}_{\ell}$ and ${\mathcal{O}}_{m}$ can be written as
follows~{\footnote{
{Evidently, when $m/N\sim 1$ the $1/N$  expansion breaks down and the  corresponding series has to be
resummed. This can be done using the semiclassical  expansion, see. refs.~\cite{Badel:2019oxl,Henriksson:2022rnm} for discussion and
references. }}
\begin{align}
\widetilde{\gamma}_{\ell}& =-\ell(\ell-2)\gamma_\pi + C_2^\ell\,\widetilde{\gamma}_{2}+
C_3^\ell\, \Delta\widetilde\gamma_{3} +O(1/n^3)\,,
\notag\\
\gamma_m &=
\frac13 {m(2m+1)}(\gamma_{\pi^2} - 2\gamma_{\pi}) + \frac83 C^m_2\left(\widetilde{\gamma}_{2} - 3\gamma_\pi\right)
+\frac{16}{5} C^m_3 \Delta\widetilde\gamma_3
\notag\\
&\quad +\frac{4}5(2m+1) C^m_2 \Delta\gamma_3 +O(1/n^3)\,,
\end{align}
where  $\Delta\gamma_3$ and $\Delta\widetilde\gamma_3$ are  the contributions to the anomalous dimensions of operators $\pi^a\pi^2$ and
$\widetilde{\mathcal{O}}_3$ coming from the diagrams listed in Fig.~\ref{fig:D13} and Fig.~\ref{fig:D47}. Notably, they give contribution
starting from the $1/n^2$ order. Thus, in order to fix the anomalous dimensions $\widetilde{\gamma}_{\ell}, {\gamma}_{m}$  with $1/n$
accuracy it is enough to calculate $\widetilde{\gamma}_{2}^{(1)}$:
\begin{align}
\widetilde{\gamma}_{2}^{(1)} &
=2/(\mu-1).
\end{align}
Here and in what follows we use the notation
\begin{align}
{\gamma}  = (\eta_1/n){\gamma}^{(1)} +
(\eta_1/n)^2{\gamma}^{(2)} + O(1/n^3),
\end{align}
etc. The operator $\mathcal O_{-}$ requires separate calculation which at leading order
is quite straightforward, so that we obtain
\begin{align}\label{gammaminus1}
\gamma_+^{(1)} &=
\frac43{(5\mu-3)(2\mu-1)}/{(\mu-1)},
\notag\\
\gamma^{(1)}_{-} &=\frac43{}(2\mu-1)(2\mu-3)(\mu-2)/(\mu-1)\,.
\end{align}
The contributing diagrams are shown in Fig.~\ref{fig:LO}.

\subsection{$1/n^2$ results}\label{sect:2n}
The $1/n^2$ order diagrams contributing to the anomalous  dimensions in question are shown in Fig.~\ref{fig:singleboxes} and
Fig.~\ref{fig:DoubleBoxes}. The diagrams in the CH and  the GN models are the same up to  isotopic factors. The values for most of the
diagrams can be retrieved from the refs.~\cite{Vasiliev:1993pi,Manashov:2017rrx}. We collected them in the~\ref{app:two-diagrams}. In
addition to the diagrams listed in Figs.~\ref{fig:singleboxes} and~\ref{fig:DoubleBoxes}, the anomalous dimensions receive contributions
from operator-vertex correction diagrams (Fig.~\ref{fig:OC}), as well as from self-energy and vertex corrections diagrams. The results of
the corresponding calculations, together with brief descriptions of the methods used, are presented in~\ref{app:ovc}
and~\ref{app:vertex-se}, respectively.

Let us introduce the following notations~\cite{Vasilev:2004yr} for the functions which appear in the final expressions:
%
\begin{align}
\alpha&\equiv \mu-1,
\notag\\
B(z) & =\psi(z)+\psi(\mu-z),
&& C_z = \psi^\prime(z)-\psi^\prime (\mu-z),
\notag\\
B_n & =B(n-\mu)-B(1),
&&\mathrm\Phi_n =C_{n-\mu}-B_n^2\,.
\end{align}
We obtain
\begin{align}
\label{three-particle-anomalous-dim}
\Delta\gamma_3^{(2)} &=-\frac{10}3\frac{\mu^2}{\alpha}\left\{\frac{5}3+2\alpha+\frac{4}{3\alpha}-\frac2{5\alpha^2}+
\left(\alpha - \frac25\right)C_1
\right\},
\notag\\
\Delta\widetilde \gamma_3^{(2)} &=-\frac{16}3 \frac{\mu^2}{\alpha}\Big\{C_1+1/{\alpha^2}\Big\}\,,
\end{align}
 where $\Delta \gamma_3=(\eta_1/n)^2 \Delta \gamma_3^{(2)} + O(1/n^3)$, etc.

For $\widetilde\gamma_2^{(2)}$ we derive
\begin{align}
\widetilde\gamma_2^{(2)} &=
\frac{\mu}{6\alpha}\Bigg\{-\frac{8}{\eta_1}\left(4 + \left( \frac{2\alpha-1}{\alpha-1}\right)^2\right)
-\frac{4\mu(3\alpha-2)}{\alpha-1}C_1
\notag\\
&\quad
+\frac43\frac{8\alpha^5+2\alpha^4-7\alpha^3-29\alpha^2+15\alpha+7}{(\alpha+1)\alpha(\alpha-1)^2} B_1
\notag\\
&\quad
+ \frac{2}3\Big[28\mu  - 54 - \dfrac{9}{\mu} + \dfrac{9}{\mu^2} - \dfrac{66}{\alpha} - \dfrac{8}{\alpha^2} + \dfrac{19}{\alpha-1}
\notag\\
&\quad
+ \dfrac{10}{(\alpha-1)^2}\Bigg]\Bigg\}.
\end{align}
Together with $\gamma_{\pi^2}^{(2)}$ \cite{Gracey:2018qba}, these functions determine the anomalous  dimensions
$\widetilde{\gamma}_\ell,\gamma_m$ with $1/n^2$ accuracy.

Finally,
 we present an expression for the anomalous dimension $\gamma_+^{(2)}\equiv\gamma_2^{(2)}$:
\begin{align}
\gamma_+^{(2)} &= \frac{2\mu}{9\alpha}\Biggl\{
-\frac{12}{\eta_1} \frac{(24\alpha^2-16\alpha+5)}{(\alpha-1)^2}-\frac{20\mu(2\alpha-1)}{\alpha-1}\mathrm \Phi_2
\notag\\
&\quad
+\mu\left[ \frac{62}{\alpha-1} -30\alpha+127
\right] C_1
+2\Big[
60\alpha^2+24 \alpha +\frac{9}{\mu}
\notag\\
&\quad
 +\frac{18}{\alpha}-\frac{65}{\alpha-1} -\frac{26}{(\alpha-1)^2}
-47
\Big] B_1 -240\mu^2  +48\mu
\notag\\
&\quad
+114 -\frac{9\alpha}{\mu^2} -\frac{61}\alpha -\frac1{\alpha^2}
+\frac{407}{\alpha-1}+\frac{130}{(\alpha-1)^2}
\Biggr\}\,.
\end{align}
For the anomalous dimension $\gamma_-$ we obtain
\begin{align}\label{gammaminus2}
\gamma_-^{(2)} &= \frac{\mu(1-2\alpha)}{9\alpha}\Biggl\{
\frac 6{\eta_1}\biggl[32\alpha^2 +76-\frac{21}\mu-\frac{3}{\alpha-1} +\frac{112}{\alpha-2}\Big]
\notag\\
&\quad
-16 \mathrm \Phi_3+\biggl[55-12\alpha^3+18\alpha^2+6\alpha+\frac{18}{2\alpha-1}\biggr]C_1
\notag\\
&\quad  +4\biggl[23-16\alpha^3
+40\alpha^2- 52\alpha
-\frac{36}\mu  +\frac{4}\alpha +\frac{13}{2(\alpha-1)}
\notag\\
&\quad -\frac{14}{\alpha-2}\biggr]B_3 +104\alpha^3-176\alpha^2 -10\alpha+196 -\frac{350}{\mu}
\notag\\
&\quad
+\frac{153}{\mu^2}-\frac{30}\alpha+\frac{17}{\alpha^2}-\frac{20}{\alpha-1}
-\frac{6}{(\alpha-1)^2}+\frac{28}{\alpha-2}\Biggr\}\,.
\end{align}
Expanding the correction exponents
\begin{align*}
\omega_\pm =2\epsilon + \gamma_\pm =\sum_k \omega_\pm^{(k)}/n^k
\end{align*}
in a power series in $\epsilon$ we obtain
 \begin{align}
    \omega_{+}^{(0)} &=2\epsilon,
\notag \\
    \omega_{+}^{(1)} &=84\epsilon - 158\epsilon^2 - 7\epsilon^3 + \left(-\dfrac{23}{2} + 168\zeta_3\right)\epsilon^4
    \notag\\
    &\quad
    -\left(\frac{55}4+316\zeta_3-252\zeta_4\right) \epsilon^5  + O(\epsilon^6),
\notag \\
    \omega_{+}^{(2)} &=- 1932\epsilon   +
    4141\epsilon^2  + \left (\frac{8935}{2} + 2352\zeta_3\right)\epsilon^3
    \nonumber \\ &- 4\bigg(1913+
    4404\zeta_3 - 882\zeta_4  + 1080\zeta_5 \bigg)\epsilon^4
    \notag\\
    &\quad
    +\Big(-1661 +30772\zeta_3-26424\zeta_4+29504\zeta_5
    \notag\\
    &\quad -2976\zeta_3^2-10800\zeta_6 \Big)\epsilon^5  + O(\epsilon^6).
\intertext{and}
    \omega_{-}^{(0)} &=2\epsilon,
\notag \\
    \omega_{-}^{(1)} &=-12 \epsilon^2+38 \epsilon^3-17 \epsilon^4 -\left(\frac{25}2+24\zeta_3\right)\epsilon^5  + O(\epsilon^6),
    \notag \\
    \omega_{-}^{(2)} &=  +
    \frac{15}2\epsilon^2  + \left (\frac{307}{4} -36\zeta_3\right)\epsilon^3
    \notag \\
    &\quad
 + \bigg(-\frac{1891}2
    +42\zeta_3 -54\zeta_4  \bigg)\epsilon^4
  \notag  \\
   &\quad
    +\left(\frac{23633}8 +887\zeta_3 + 63\zeta_4 -1232\zeta_5\right)\epsilon^5  + O(\epsilon^6)\,.
\end{align}
Both indices are in complete agreement with the $\epsilon$ expansion near $d = 4$, which is known with $\epsilon^4$ accuracy~\cite{Zerf:2017zqi}.

\section{$d\to 3$ limit}\label{sect:dto3}
The large $N$ expansion provides not only a check for  perturbative calculations but also allows one to access the indices in the
dimension~$d=3$. Unfortunately,  almost all  physical systems  of interest  do not have a large symmetry group, $N\sim 1$. For small $N$
the higher-order corrections are quite sizeable, e.g.
\begin{align}
\eta  &= \frac1n \frac{4}{\pi^2} \left(1+\frac 1{n} \frac{16}{3\pi^2}\right) + O(1/n^3)\,,
\notag\\[2mm]
\gamma_\pi &=\frac{1}{n^2}\frac{8}{\pi^2} \left(1+\frac{16}{3\pi^2}\right)+ O(1/n^3)\,,
\end{align}
and grow  with the dimension of an operator~\footnote{We recall that $n=N\times \text{tr} \mathds{1} =2 N$  in $d=3$ }
\begin{align}
2\lambda &= -\frac1n \frac{16}{\pi^2}\left(1-\frac1{n}\left(3+\frac{104}{3\pi^2}\right)\right)+ O(1/n^3)\,,
\notag\\
\widetilde \gamma_2 &= \frac1n \frac{16}{\pi^2}\left(1-\frac1{n}\left(3-\frac{40}{3\pi^2}\right)\right)+ O(1/n^3)\,,
\notag\\
\gamma_+ &= \frac1n \frac{96}{\pi^2}\left(1-\frac1{n}\left(\frac{10}3 + \frac{488}{9\pi^2}\right)\right)+ O(1/n^3)\,.
\end{align}
Moreover, at $d=3$ the anomalous dimension $\gamma_-$ has a pole~\footnote{ The pole arises exclusively from the term $\Phi_3$ in
Eq.~\eqref{gammaminus2}.}
\begin{align}
\label{gamma11-res}
\gamma_-(d) & \underset{d\sim 3} {=} -\left(\frac{\eta_1}{n}\right)^2\frac{32}{3(d-3)} + O(1)\,.
\end{align}
The appearance of a pole usually indicates that the  perturbative series has to be resummed. As a rule, the singularities  are associated
with the operator mixing, see ref.~\cite{Korchemsky:2015cyx}  for general analysis.
 One can distinguish two different scenarios.

\vskip 2mm \noindent
 I.~ Consider two operators $\mathcal{O}_1$ and $\mathcal{O}_2$
which have the same canonical dimension, $\Delta_{\text{can}}$, for arbitrary $d$. Such operators mix in the $1/N$ expansion.  Their
scaling dimensions are determined by the eigenvalues of the anomalous dimension matrix $\gamma_{ik}$ and take the form
\begin{align}\label{Deltaik}
\Delta_{i}=\Delta_{\text{can}} + \frac12\left(\gamma_{11}+\gamma_{22}\pm\sqrt{(\gamma_{11}-\gamma_{22})^2+4\gamma_{12}\gamma_{21}} \right).
\end{align}
Let us assume that the diagonal elements $\gamma_{kk}\sim 1/n$ and the product $\gamma_{12}\gamma_{21}\sim 1/n^3$. Assuming also that
$\gamma_{11}^{(1)}\neq \gamma_{22}^{(1)}$ and expanding $\Delta_i$ in $1/n$ one gets
\begin{align}\label{Delta1N}
\Delta_{i} &= \Delta_{\text{can}} +  \gamma_{ii}\pm
\dfrac{n}{\eta_1}\dfrac{\gamma_{12}\gamma_{21}}{\gamma_{11}^{(1)}-\gamma_{22}^{(1)}} +O(1/n^3).
\end{align}
Thus if at some point $d = d_0$ our assumption is violated and leading-order anomalous dimensions intersect
$$\gamma_{11}^{(1)}(d)-\gamma_{22}^{(1)}(d)\sim c(d-d_0),$$
the perturbative expansion~\eqref{Delta1N} diverges. At the same time the {\it exact} scaling dimensions, Eq.~\eqref{Deltaik},  do not have
poles but they do have branching points, $d_\pm =d_0 + O(1/\sqrt{n})$. The concrete examples can be found in
refs.~\cite{Derkachov:1998js,Korchemsky:2015cyx}. Remarkably, the knowledge of the anomalous dimensions, Eq.~\eqref{Delta1N}, outside the
intersection point, makes it possible to restore the correct behavior at the point $d = d_0$ by solving a quadratic equation
\begin{equation}\label{AlG}
(\Delta-\Delta_1)(\Delta-\Delta_2) =\Delta^2 -\Delta (\Delta_1+\Delta_2) +\Delta_1\Delta_2=0.
\end{equation}
While the exponents $\Delta_k$ have poles, their sum  $\Delta_1 + \Delta_2$ and the product $\Delta_1\Delta_2$ are finite at $d= d_0$ up to
$1/n^3$ and $1/n^4$, respectively, It allows one to get the finite solutions at the intersection point.

\vskip 5mm \noindent II.~ A somewhat different situation arises when considering operators with different canonical dimensions,
$\Delta_k(d)$, for an arbitrary $d$, which coincide however for some value of $d$. An example of such a system is:
$$
A=\{\pi\partial^2\pi, \pi^4 \}
$$
 and
$$
B=\{(\bar{q}q)^2, (\bar{q}\sigma^aq)^2, (\bar{q}\gamma_{\mu}q)^2, \ldots\},
$$
the system of all possible four-fermion operators.
The canonical dimensions of the  operators in the $A$ and $B$ groups are different:
\begin{align}
\label{ab-scaling}
\Delta_A=4, && \text{and} &&\Delta_B =2d-2.
\end{align}
However,  $\Delta_A=\Delta_B$  for  $d=3$. It means that the diagrams which are finite for $d\neq 3$ start to diverge as $d=3$. This
already happens at leading order in $1/N$ and results in
\begin{align}
\lim_{d\to 3}\gamma_k(d) \neq \gamma_k(3)\,.
\end{align}
The situation strongly resembles that observed in ref.~\cite{Giombi:2017rhm} where it was noticed that in the GN model the anomalous
dimension of the operator $\sigma \partial_+^s\sigma$ in the limit $s\to0$ differs from the anomalous dimension of the
operator $\sigma^2$ at leading order in $1/N$.

\vskip 3mm

Let us now explore the mixing in $d=3$ in more details. The operator $\mathcal O_1=\pi\partial^2\pi$ mixes with the  four-fermion
operators~\footnote{There is no mixing with the operator $(\bar q q)^2$ at order $1/n$. We also omit the mixing with the EOM operator
$(\bar q\sigma^a q)^2$.}
\begin{align}\label{O123def}
\mathcal O_2 = (\bar{q}\sigma^a\gamma_{\mu}q)^2,
&&
\mathcal O_3= (\bar{q}\gamma_{\mu\nu}q)^2 \sim (\bar{q}\gamma_{\mu}q)^2,
\end{align}
where  $\gamma_{\mu\nu} = \tfrac{1}{2}\left[\gamma_{\mu}, \gamma_{\nu}\right]$. The anomalous
dimension matrix at order $1/n$ takes the form:
\begin{align}
\label{3d-matrix}
\gamma=\frac{\eta_1}{n}
\begin{pmatrix}
0 & -\frac{512 n^2}3 & 0\\
-\frac1{8n^2} & 0&-4\\
0 & -\frac{16}3 & 0
\end{pmatrix}+O(1/n^2)\,.
\end{align}
Note that operators $\bar{q}\gamma_{\mu}q$ and $\bar{q}\sigma^a\gamma_{\mu}q$ are the conserved currents.  The eigenvalues of the matrix
are: $\gamma_0 = 0 + O(1/n^2)$, and
 \begin{align}
 \label{d3-result}
\gamma_\pm =\pm \frac{8\eta_1}{n}  \sqrt{\frac 23} +O(1/n^2).
\end{align}
\vskip 3mm

After discussion of mixing exactly in $d = 3$, let us see what happens if one computes the matrix of anomalous dimension in $d > 3$ and
then takes the limit $d\to 3$. The first difference with $d=3$ case   is that the system of four-fermion operators $B$ becomes infinite. It
 consists of operators of the following type
\begin{align}
\mathcal O^+_k &=
\Big(\bar q\gamma_{\mu_1\ldots\mu_k} q\Big)\Big(\bar q\gamma^{\mu_1\ldots\mu_k} q\Big),
\notag\\
{\mathcal O}^-_k &=
\Big(\bar q\sigma^a\gamma_{\mu_1\ldots\mu_k} q\Big)\Big(\bar q\sigma^a \gamma^{\mu_1\ldots\mu_k} q\Big)\,,
\end{align}
where $\gamma_{\mu_1\ldots\mu_k}$ represents the fully antisymmetric combination of $\gamma$-matrices. Fortunately, at order $1/n$ the
mixing is block diagonal: the operator $(\bar q q)^2$ is multiplicatively renormalizable, and for each $k=0,2,4,\ldots$, the following
operators mix
\begin{align}
B_k=\Big\{\mathcal O^+_{k-1},{\mathcal O}^-_{k},{\mathcal O}^-_{k+1},\mathcal O^+_{k+2} \Big\}\,,
\end{align}
where the first block, $B_0$, that we are mostly interested in, contains only  three operators, one of which, $\mathcal O^-_0$, is an EOM
operator. All blocks have the same eigenvalues~\cite{Ji:2018emi} thus the anomalous dimensions are infinitely degenerate at leading
order.

This degeneracy is lifted by higher order corrections. However,  as we show in what follows, only divergencies of anomalous dimensions in
the next order can contribute to the leading-order result. At order $1/n^2$ it is sufficient to consider mixing inside the block and we
argue that for the blocks $B_{k\geq 2}$ the corrections do not have poles at $d=3$. The only divergencies appear in the block $B_0$. This
can be justified by considering the Feynman diagrams contributing to the corresponding anomalous dimensions. The diagram giving a singular
contribution at $d = 3$ to the anomalous dimension $\gamma_{22}$ is shown in Fig.~\ref{fig:4f} (plus other diagrams with different
orientation of the fermion lines). For non-integer $d$ this diagram has only superficial divergence. However, the upper subgraph $D_1$ in
the dashed box starts to diverge at $d=3$ as well as the diagram $D_2$ arising after contraction of this subgraph. The upper subgraph,
$D_1$, has a pole in the parameter $\omega=d-3-\Delta$, and the subgraph $D_2$ in $\omega'=d-3+3\Delta$,  for a general discussion see
ref.~\cite{Speer:1969sjv},
\begin{align*}
D_1\sim \frac{1}{d-3-\Delta}, && D_2\sim \frac{1}{d-3+3\Delta}.
\end{align*}
Thus, one sees from this example that the poles in $\Delta$ for $d=3$ and the poles in $d-3$ for $\Delta=0$ are closely related.
Calculating the corresponding diagrams we find the singular (at $d\to 3$) contribution  to the matrix element $\gamma_{22}$,
\begin{align}
\label{gamma22-res}
\gamma_{22}  \underset{d\sim 3} {=} \left(\frac{\eta_1}{n}\right)^2\frac{32}{3(d-3)} + O(1)\,.
\end{align}

\begin{figure}[t]
  \centering
  \includegraphics[width=0.5\columnwidth]{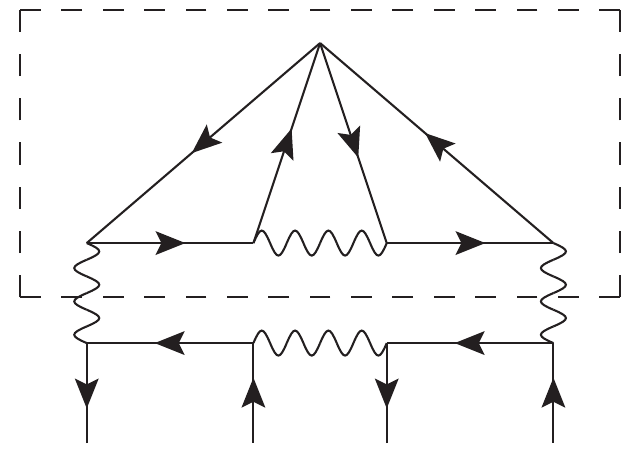}
  \caption{ An example of a diagram contributing to the $d-3$ pole in the anomalous dimension $\gamma_{22}$.
  }
  \label{fig:4f}
\end{figure}

The pole at $d=3$ in the anomalous dimensions $\gamma_{11}$, $\gamma_{22}$, Eqs.~\eqref{gamma22-res},~\eqref{gamma11-res},  indicates that
the $1/n$ series has to be resummed in all orders in $1/n$ to give a meaningful result in the limit $d \to 3$. Following the lines of
ref.~\cite{Korchemsky:2015cyx} and  taking in mind the examples discussed at the beginning of this section as well as advances in the
studies of the analytical structure of scaling dimensions in spin~\cite{Caron-Huot:2022eqs} we present such  a resummation procedure. More
examples can be found
 in~\cite{Korchemsky:2015cyx,Velizhanin:2011pb,Velizhanin:2014dia,Velizhanin:2022seo,Caron-Huot:2022eqs,Manashov:2025kgf}.

Let us consider the  operators $\mathcal O_1, \mathcal O_2,\mathcal O_3$, see Eq.~\eqref{O123def}. In arbitrary $d$, the first operator
does not mix with the other two, and their scaling dimensions are determined by the matrix
\begin{equation}\label{gamma1n}
    \widehat{\Delta} = 2d - 2 + \underbrace{\begin{pmatrix}
        \gamma_{11}(d) - 2(d - 3) & 0 & 0 \\
        0 & \gamma_{22}(d) & \gamma_{23}(d) \\
        0 & \gamma_{32}(d) & \gamma_{33}(d)
    \end{pmatrix}}_{M},
\end{equation}
where $\gamma_{11}\equiv \gamma_-$. The lower $2\times 2$ block at order $1/n$ takes the form
\begin{align}
\label{lo-fermion}
\gamma_{2\times 2} =\frac{\eta_1}n \frac4{d-2}\begin{pmatrix}
0 & - 1 \\
-\frac23(d-1) & d-3
\end{pmatrix} +O(1/n^2)\,.
\end{align}
 Away from $d = 3$, the anomalous dimensions are obtained by solving the characteristic equation for the matrix $M$
\begin{equation}
\label{char-eq}
    P(\lambda) = \det(M - \lambda\mathds{1}) = 0,
\end{equation}
yielding three eigen-trajectories:  one corresponding to the operator $\pi\partial^2\pi$, $\Delta_{-}(d) = 4 + \gamma_{-}(d)$, and two
others corresponding to the four-fermion operators. At $d \sim 3$, the $1/n$ corrections become large and the expansion loses its meaning.
Nevertheless, keeping in mind the example from the beginning of this section, we rewrite~\eqref{char-eq} in the form
\begin{equation}\label{Plambda}
    P(\lambda) = -\lambda^3 + A(d)\lambda^2 - B(d)\lambda + C(d) = 0,
\end{equation}
where
\begin{align}
\label{char-eq-coefficients}
	A(d)&= 2(3 - d) + \gamma_{11} + \gamma_{22} + \gamma_{33},
\nonumber\\
        B(d) &= \big(2(3 - d)+\gamma_{11}\big)\big(\gamma_{22} + \gamma_{33}\big)
	+ \gamma_{22}\gamma_{33} - \gamma_{23}\gamma_{32},
\nonumber\\
        C(d) &= \big(2(3 - d)+\gamma_{11}\big)\big(\gamma_{22}\gamma_{33} - \gamma_{23}\gamma_{32}\big).
\end{align}
Taking in mind the example discussed earlier in this section, see   Eq.~\eqref{AlG}, we expect that while the anomalous dimensions
$\gamma_{ik}$ have poles at $d=3$ the coefficients $A, B, C$ are regular functions at this point. Taking into account
Eqs.~\eqref{gamma11-res},~\eqref{gamma22-res} and \eqref{gamma1n} it can be checked that this indeed the case: the coefficient $A(d)$ is
free of poles up to $O(1/n^3)$ terms; $B(d), C(d)$ --  up to $O(1/n^4)$ terms. It is natural to expect that $C(d)$ is finite at order
$1/n^4$, but in order to check this we need to know the elements $\gamma_{22},\ldots,\gamma_{33}$ at order $1/n^3$. One should also take
into account that at higher orders Eq.~\eqref{Plambda} could be modified  due to mixing with  blocks, $B_{k>0}$.

\vskip 2mm

Let us find the solutions of Eq.~\eqref{char-eq} for $d=3$ at leading order in $1/n$. In $d=3$ the
coefficients~\eqref{char-eq-coefficients} take the form
\begin{align}
    A(3) &=0 + O(1/n^2), \nonumber \\
    B(3) &= -\dfrac{128}{3}\left(\dfrac{\eta_1}{n}\right)^2 + O(1/n^3),
    \nonumber \\
    C(3) &= 0+O(1/n^4).
\end{align}
Thus, Eq.~\eqref{char-eq} at $d = 3$ at leading order in $1/n$ is reduced to
\begin{equation}
    \lambda\left(\lambda^2 - \dfrac{128}{3}\left(\dfrac{\eta_1}{n}\right)^2\right) = 0,
\end{equation}
which  gives~$\gamma_0,\gamma_\pm$, Eq.~\eqref{d3-result}, as solutions.

\begin{figure}[t]
  \centering
  \includegraphics[width=0.9\columnwidth]{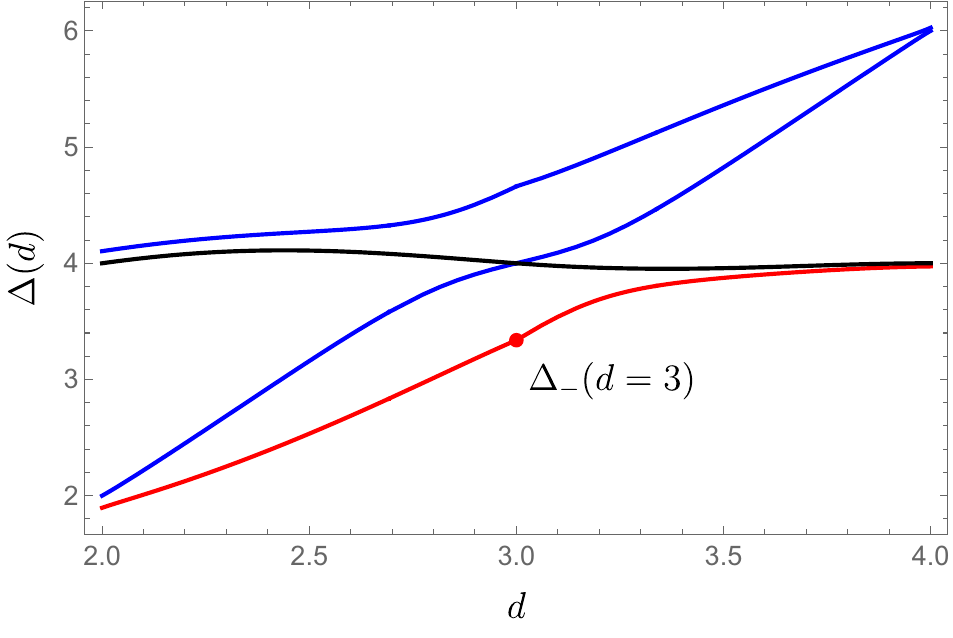}
  \caption{Comparison of the resummed and naive scaling dimensions in $1/n$ approximation.
  The blue and red lines correspond to the solution of~\eqref{char-eq},
  while the black line represents the "naive" scaling dimension~\eqref{gammaminus1}.
   The graph is made for $N = 2$ and $\operatorname{tr}\mathds{1}$ is approximated as $2$ for $2 < d < 3$ and $2d - 4$ for $3< d < 4$.
   }
    \label{fig:trajectories}
\end{figure}

We  also plot the solutions of Eq.~\eqref{Plambda} (see Fig.~\ref{fig:trajectories}) in the region $2 < d < 4$ by calculating the
leading-order coefficients~\eqref{char-eq-coefficients}. Here we use the results~\eqref{lo-fermion} and~\eqref{gammaminus1} at order $1/n$
and take into account only contributions divergent at $d = 3$ at order $1/n^2$, Eqs.~\eqref{gamma11-res} and~\eqref{gamma22-res}. When $d$
is far from the point $d = 3$, we observe that solutions reproduce the "naive" result for the anomalous dimension of the operator
$\pi\partial^2\pi$,~\eqref{gammaminus1}, quite well. However, as $d$ approaches $3$, the deviation becomes significant. The scaling
dimension $\Delta_{-}$ is associated with the solution, which approaches $\Delta = 4$ at $d \to 4$ (red line in
Fig.~\ref{fig:trajectories}). One sees that the $1/n$ order result changes~\footnote{We recall that $\eta_1\big|_{d=3}=4/\pi^2$.}
\begin{align}
0 \rightarrow -\frac{8\eta_1}{n}\sqrt{\dfrac{2}{3}}.
\end{align}
The resummed scaling dimensions of four-fermion operators also deviate from their ``naive" values:
 \begin{align}
-\frac{8\eta_1}{n}\dfrac{1}{\sqrt{3}} \longrightarrow 0  &&\text{and}&& \frac{8\eta_1}{n}\dfrac{1}{\sqrt{3}}
\longrightarrow \frac{8\eta_1}{n}\sqrt{\dfrac{2}{3}}.
\end{align}
The  deviation from the "naive" leading-order trajectories is controlled by the parameter $\delta= 1/n(d-3)$: for  $\delta \ll 1$
deviations are small.
Note also that at $d=5/2$ and $d=7/3$  there will be another "crossing" points  where   the trajectory corresponding to the operator
$\pi\partial^2\pi$ intersects  the trajectories of four-fermion operators with one derivative and six-fermion operators, respectively.

Finally,  we note that in the GN model the $d=3$ pole contributions to the anomalous dimension $\gamma_-$, coming from the diagrams,
$DB_2-DB_4$, shown in Fig.~\ref{fig:DoubleBoxes} cancel out. It means that the anomalous dimension of the operator $\sigma\partial^2\sigma$
remains unchanged at order $1/n$. This agrees with the fact  that there is no mixing of the operator $\sigma\partial^2\sigma$ and
four-fermion operator $(\bar q \gamma_\mu q)^2$ in $d=3$ at leading order.

\section{Summary}\label{sect:summary}
We calculated  the correction exponents $\omega_\pm$, related to the slopes of the $\beta$ functions, at order $1/n^2$ in the chiral
Heisenberg model in general dimension $d$. Expanding the exponents $\omega_\pm$ in an $\epsilon$-series near four dimensions we found full
agreement with the results of the perturbative calculations~\cite{Zerf:2017zqi}. We also found that the index $\omega_-(d)$ at order
$1/n^2$ has a pole at $d=3$. This happens because divergent subgraphs appear in some diagrams (namely, double boxes in
Fig.~\ref{fig:DoubleBoxes}) at $d=3$. This situation is quite common. The canonical dimensions of operators depend on the space-time
dimension $d$ and for certain values,
 $d=d_*$, the number of mixing operators changes. This means that some diagrams that are finite for general $d$ start to diverge at
$d_*$,
$$
D\sim \frac A{\#(d-d_*)+\Delta}\,.
$$
Indeed, for $d\neq d_*$ such a diagram is finite, but has a pole at $d=d_*$, whereas for $d=d_*$ the diagram diverges and contributes to
the renormalization factor. Since GN-like fermion models are renormalizable at $d=3$ in the large $N$ expansion the critical indices of the
fermion and the auxiliary fields are continuous at $d=3$ at any order in $1/N$. The same is true for the low-dimensional operators,
$\Delta_O\leq 3$, whose scaling dimensions should be regular at $d=3$. However, for the critical exponents of higher-dimensional operators,
$\Delta_Q\geq 4$, the scenario described above is quite possible, and one should expect that $\gamma_{ik}(d)$ diverges as $d\to 3$.

{
{ It should be noted that the observed effect is not only a feature of the model under consideration, but is also present
in bosonic models such as the nonlinear sigma model.  }}

 Thus, obtaining
the critical exponent for operators of high dimensions, $\Delta_Q\geq 4$, is quite subtle. Although the divergencies appear only at
next-to-leading order, they affect the critical exponents already at leading order. Of course, it is always  possible  to perform all
calculations in the three-dimensional model. However, in this way one loses all connection with the $4-2\epsilon$ expansion that is highly
undesirable. Remaining within the framework of the general $d$ formalism, it is necessary to resum the individual contributions. Taking
into account recent advances in understanding the analytical structure of the anomalous dimensions of leading twist operators at small
conformal spins, see
refs.~\cite{Velizhanin:2014dia,Korchemsky:2015cyx,Caron-Huot:2022eqs,Manashov:2025kgf},
we proposed a resummation formula and found complete agreement between the leading-order indices calculated in both approaches.

Numerically, the deviation of the resummed exponent from its "naive" value is quite significant. It would also be interesting to test
whether this effect is present in the $\epsilon$ expansion, as was observed in the critical Ising model~\cite{Henriksson:2022gpa}.

\begin{acknowledgements}
We are grateful to John Gracey for attracting our attention to this problem and Sergey Derkachov, Gregory Korchemsky and Sven-Olaf Moch for helpful
discussions and critical remarks. The work of A.N.M. has been supported by the Deutsche Forschungs-gemeinschaft (DFG) Research Unit FOR
2926, “Next Generation pQCD for Hadron Structure: Preparing for the EIC”, project number 40824754, and the work of L.A.S. has been
supported by the ERC Advanced Grant 101095857 {\it Conformal-EIC}.
\end{acknowledgements}

\appendix

\section{Results for the individual diagrams\label{app:two-diagrams}}

In this appendix we collect the results for the individual Feynman diagrams we use in our analysis. For each diagram we present a
 divergent part after subtraction of all subdivergencies. The contribution from the diagram $D$
 to the corresponding anomalous dimension  reads
\begin{align}
    \Delta\gamma_D &= -2\partial_u D_1 \Big|_{u = 1}, 
\end{align}
where $D_1$ is a simple pole in the diagram after subtraction of subdivergencies
$$
KR^\prime(D)=\frac1 {\Delta^2}D_{2}+\frac{1}{\Delta}D_1\,.
$$

\subsection{ $2\to 2$ diagrams}

\begin{figure}[t]
  \centering
  \includegraphics[width=0.99\columnwidth]{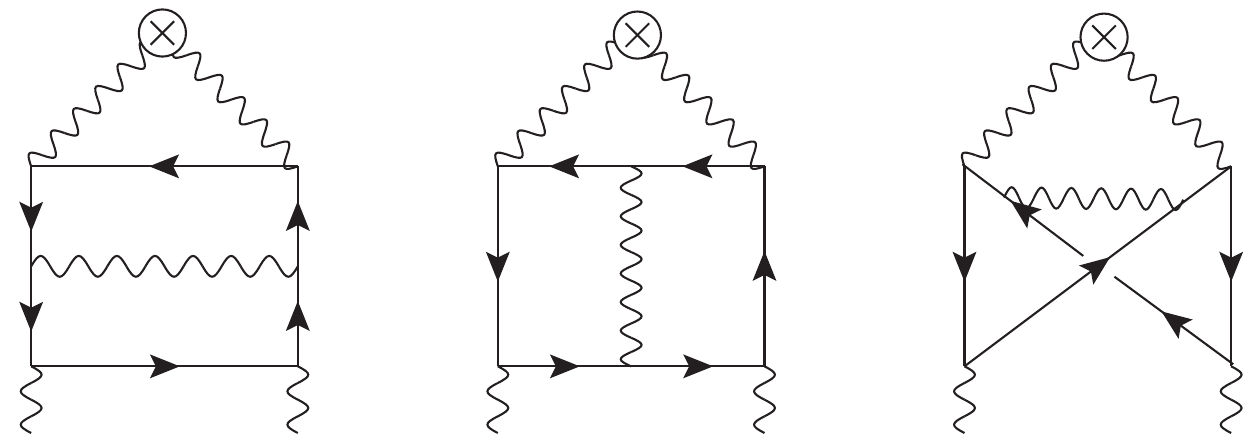}
  \caption{The simple box (SB)  diagrams.
   }
    \label{fig:singleboxes}
\end{figure}
Here we list diagrams presented in Fig.~\ref{fig:singleboxes}~and~\ref{fig:DoubleBoxes}. The form of diagrams is the same  for operators
$\pi^2$, $\pi^{\{a}\pi^{b\}} $ and $\pi\partial^2\pi$.
 Expressions for the operators $\pi^2$ and $\pi^{\{a}\pi^{b\}}$ coincide up to symmetry factors.

\begin{figure}[t]
  \centering
  \includegraphics[width=0.98\columnwidth]{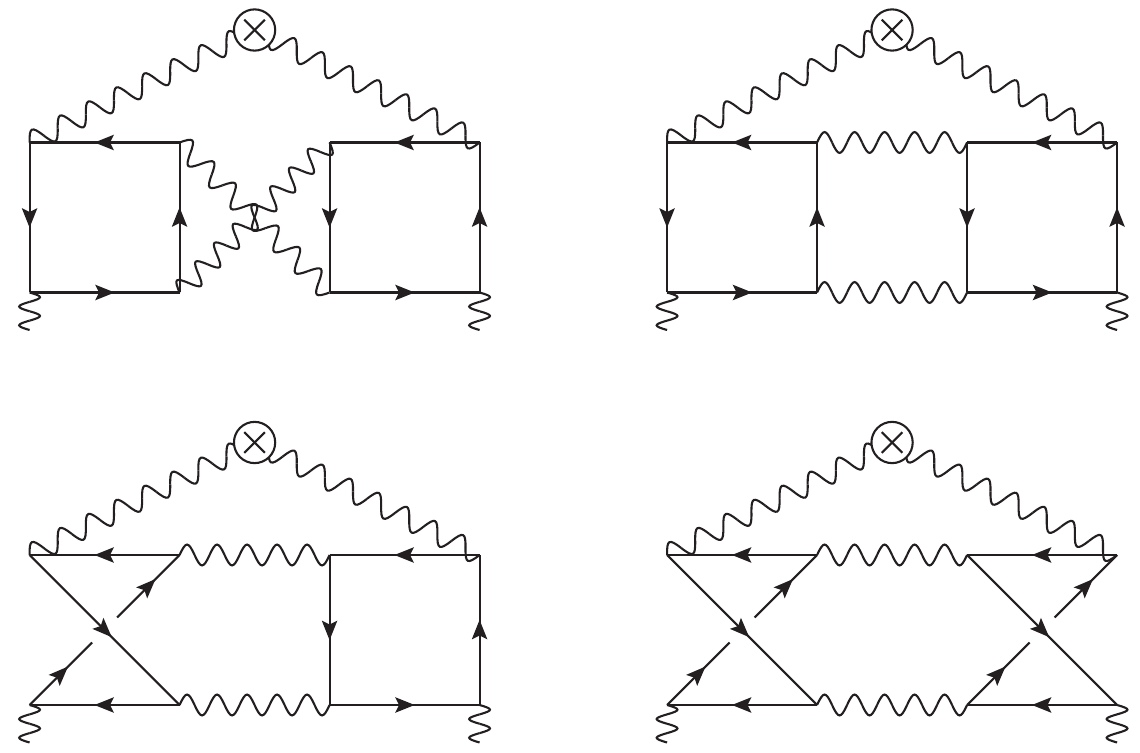}
  \caption{ Double box diagrams $DB_1$ (top left) - $DB_4$ (bottom right).
}
  \label{fig:DoubleBoxes}
\end{figure}

\begin{itemize}
\item Operators $\pi^2$ and $\pi^{\{a}\pi^{b\}} $

For the single box diagrams we get:
\begin{align}
    SB^{+}_1 &= 
     \left[18, 0\right]
     \dfrac{u^3\eta_1^2}{54\Delta}{\mu^2}/{\alpha},
\notag \\
    SB^{+}_2 &= 
     \left[2, 8\right]\dfrac{u^3\eta_1^2}{54\Delta}{\mu^2}/{\alpha^3},
\notag \\
    SB^{+}_3 &= 
     \left[10, 4\right]\dfrac{u^3\eta_1^2}{54\Delta}\dfrac{\mu^2}{\alpha^3}\left(1 + \dfrac{3}{2}C_1\alpha^2\right).
\end{align}
And for the double box diagrams we obtain:
\begin{align}
    DB^{+}_1 &= 
     \left[-4, 20\right]\dfrac{u^4}{24\Delta}\dfrac{\mu}{\alpha}\eta_1\Bigg\{1 - \dfrac{\eta_1\mu}{3\alpha}\Bigg[
      \dfrac{1}{\alpha}
 \notag \\
 &\quad +\frac{2\mu-1}2
B_2 - 1\Bigg]\Bigg\},
\notag \\
    DB^{+}_2 &= \left[28, 4\right]\dfrac{u^4\eta_1^2}{144\Delta}\dfrac{\mu^2}{\alpha}\Bigg\{\dfrac{2}{\alpha}
    - \dfrac{2}{\alpha^2} - \dfrac{6}{\alpha - 1}
 \notag \\
 &\quad
 +\frac{(2\mu-3)}{2(\alpha-1)} \big[\Phi_2-7C_1\big]
 + \left[2 
 + \dfrac{1}{\alpha(\alpha - 1)}\right]B_2
\Bigg\},
 \notag \\
    DB^{+}_3 &= \left[-4, -4\right]\dfrac{u^4}{24\Delta}\dfrac{\mu^2}{\alpha}\eta_1\Bigg\{-\dfrac{1}{\mu} - \dfrac{1}{\alpha - 1}
\notag \\
&\quad
	+ \dfrac{2\eta_1}{3}\Bigg[\dfrac{1}{\alpha} - \dfrac{1}{\alpha - 1} +\left(\dfrac{1}{2} + \dfrac{1}{4(\alpha - 1)}\right)\Phi_2
\notag\\
	&\quad
	- \dfrac{1}{2\alpha}B_2 - \left(2 + \dfrac{7}{4(\alpha - 1)}\right)C_1\Bigg]\Bigg\},
\notag \\
    DB^{+}_4 &= \left[28, 4\right]\dfrac{u^4}{96\Delta}\dfrac{\mu^2}{\alpha}\eta_1\Bigg\{\dfrac{1}{\mu} + \dfrac{3}{\alpha - 1}
    + \dfrac{2}{(\alpha - 1)^2}
    \notag \\
    &\quad + \dfrac{2\eta_1}{3}\Bigg[- \dfrac{1}{(\alpha - 1)^2} +\frac{(2\mu-3)}{2(\alpha - 1)}\left(\Phi_2-C_1\right)
\notag \\
    &\quad   + 
    \left(\dfrac{1}{\alpha - 1} + \dfrac{1}{(\alpha - 1)^2}\right) B_2 \Bigg]\Bigg\}.
\end{align}
Here $[a, b]$ represents factors coming from the traces of $\sigma$-matrices, where the first factor, $a$, corresponds to the operator
$\pi^2$ and the second one, $b$, to
the operator $\pi^{\{a}\pi^{b\}}$ 

\item Operator $\pi\partial^2\pi$.

\begin{figure}[t]
\centering
  \includegraphics[width=0.99\columnwidth]{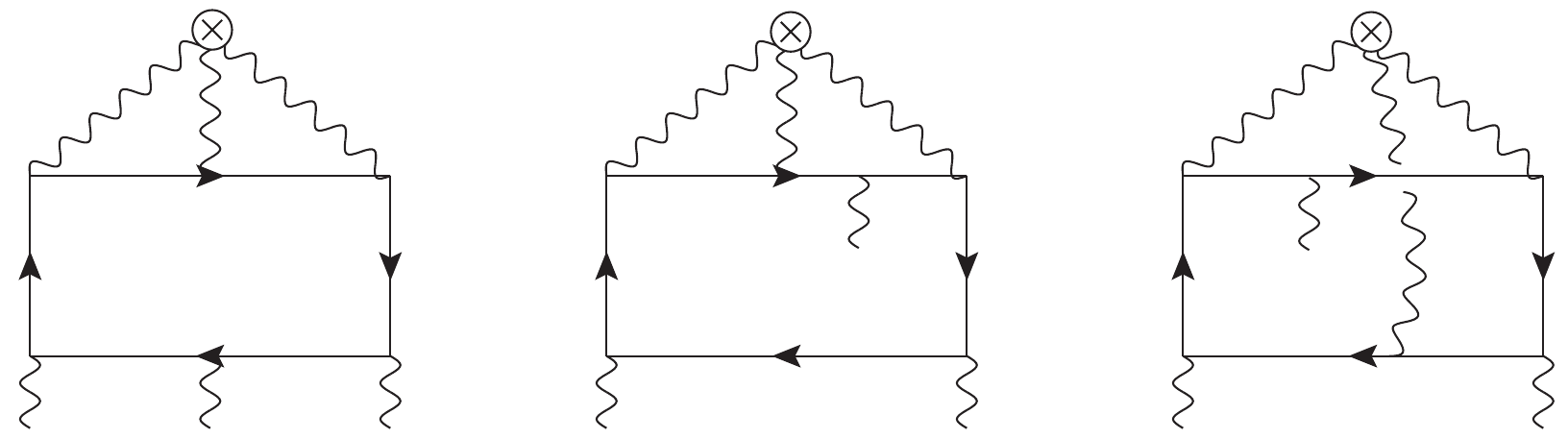}
  \caption{Diagrams $D_{1}- D_3$} 
  \label{fig:D13}
\end{figure}

Results for the single-box diagrams read:
\begin{align}
    SB^{-}_1 &= \left[18\right]\dfrac{u^3\eta_1^2}{216\Delta}\dfrac{\mu\alpha\left(2\alpha - 1\right)}{\alpha - 1},
\notag \\
    SB^{-}_2 &=  \left[2\right]\dfrac{u^3\eta_1^2}{54\Delta}\dfrac{\mu(2\alpha - 1)}{\alpha(\alpha - 1)}\Bigg(1
	- (\alpha - 1)^2 + \dfrac{1}{\alpha^2}\Bigg),
\notag \\
    SB^{-}_3 &=  \left[10\right]\dfrac{u^3\eta_1^2}{36\Delta}\dfrac{\mu}{\alpha}\Bigg(\left(\alpha^2- \alpha + 1\right)C_1
\notag \\
&\quad
	+ \dfrac{2}{3}\left(4 - 7\mu + 2\mu^2 - \dfrac{1}{\alpha} + \dfrac{1}{\alpha^2}\right)\Bigg).
\end{align}
And for the double boxes we get:
\begin{align}
    DB^{-}_1 &= \left[-4\right]\dfrac{u^4}{24\Delta}\dfrac{\mu(2\alpha - 1)}{\alpha(\alpha - 1)}\eta_1\Bigg\{2\mu^3 - 12\mu^2
    \notag \\
    &\quad
    + 25\mu + \dfrac{10}{\mu} - 24 + \dfrac{2\eta_1}{3}\Bigg[-\mu^2 + \dfrac{19\mu}{4} + \dfrac{1}{\alpha}
    \notag \\
    &\quad - \dfrac{1}{2\alpha^2} - \dfrac{1}{2(\alpha - 1)} - \dfrac{1}{4(2\mu - 3)} - \dfrac{23}{4}
    \notag \\
    &\quad + \left(\dfrac{\mu^2}{2} - \dfrac{11\mu}{4} - \dfrac{1}{4\alpha} + \dfrac{13}{4}\right)B_3\Bigg]\Bigg\},
    \notag \\
DB^{-}_2 &= \left[28\right]\dfrac{u^4}{72\Delta}\eta_1^2\dfrac{\mu(2\alpha -1)}{\alpha}\Bigg\{\alpha - \dfrac{1}{\alpha}
    \notag \\
    &\quad + \dfrac{1}{\alpha^2} + \dfrac{1}{2\mu - 3} - \dfrac{1}{4}\Phi_3 + \dfrac{7}{4}C_1
    \notag \\
    &\quad + \left(\dfrac{1}{2} - \mu + \dfrac{1}{2\alpha} + \dfrac{1}{2(\alpha - 1)}\right) B_3\Bigg\},
    \notag \\
    DB^{-}_3 &=  \left[-4\right]\dfrac{u^4}{24\Delta}\mu\alpha(2\alpha - 1)\eta_1\Bigg\{\dfrac{5}{\mu} - 4- \dfrac{1}{\alpha - 1}
    \notag \\
    &\quad
     - \dfrac{\eta_1}{3\alpha^2}\Bigg[2\mu - 1 - \dfrac{2}{\alpha} + \dfrac{2}{\alpha - 1} + \dfrac{1}{(\alpha - 1)^2} - \dfrac{2}{2\mu - 3}
    \notag \\
    &\quad 
    + \dfrac{1}{2}\left(C_1 - \Phi_3\right)
     + \left(\dfrac{1}{\alpha} + \dfrac{2}{\alpha - 1}\right)B_3\Bigg]\Bigg\},
    \notag \\
    DB^{-}_4 &=  \left[28\right]\dfrac{u^4}{48\Delta}\mu\alpha(2\alpha - 1)\eta_1\Bigg\{4 - \dfrac{3}{\mu} - \dfrac{1}{\alpha - 1}
    \notag \\
    &\quad + \dfrac{4}{\alpha - 2} + \dfrac{\eta_1}{3\alpha^2}\Bigg[3 - 2\mu - \dfrac{2}{\alpha - 1} - \dfrac{1}{(\alpha - 1)^2}
    \notag \\
    &\quad
    \!+\! \dfrac{2}{\alpha - 2} + \dfrac{1}{2}\left(C_1 - \Phi_3\right)
    -\frac{4(\alpha-1)}{\alpha-2}B_3\Bigg]\Bigg\}.
\end{align}
\end{itemize}


%

\subsection{$3\mapsto 3$ diagrams}

\begin{figure}[t]
\centering
  \includegraphics[width=0.99\linewidth]{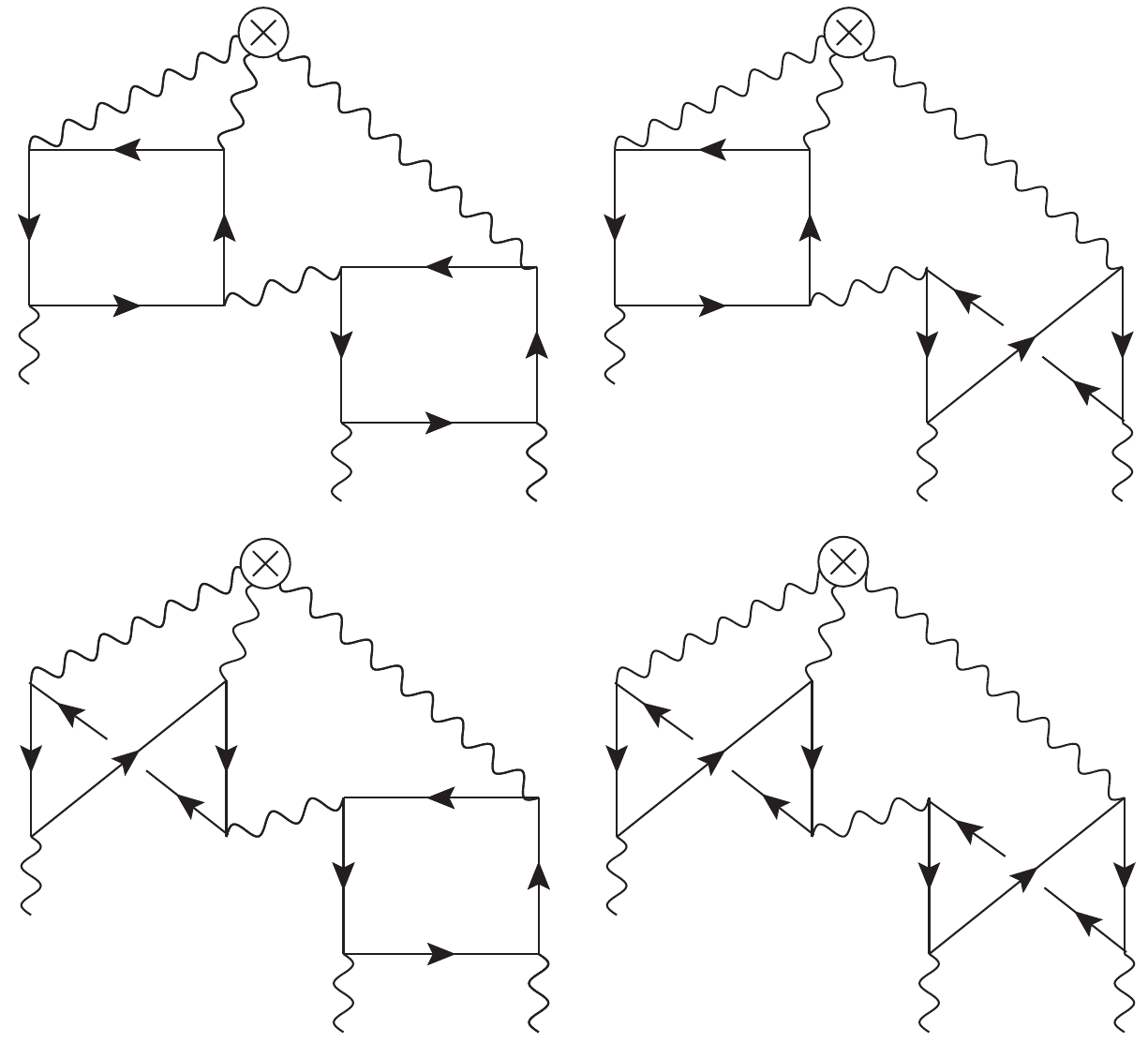}
  \caption{Diagrams $D_{4}- D_7$} 
  \label{fig:D47}
\end{figure}

Here we present results for the diagrams (see Figs.~\ref{fig:D13} and \ref{fig:D47})  contributing to the  anomalous dimensions
$\Delta\gamma_3$ and $\Delta\widetilde{\gamma}_3$, see~\eqref{three-particle-anomalous-dim}. Results for the diagrams coincide up to
symmetry factors and read:
\begin{align}
    D_1 &= -\left[0, \dfrac{10}{3}\right]\dfrac{u^3}{18\Delta}\eta_1^2\dfrac{\mu^2(2\mu - 3)}{\alpha},
\notag \\
    D_2 &= -\left[0, \dfrac{10}{3}\right]\dfrac{u^3}{6\Delta}\eta_1^2\mu^2C_1,
\notag \\
    D_3 &= -\left[8, -2\right]\dfrac{u^3}{18\Delta}\eta_1^2\dfrac{\mu^2}{\alpha}C_1,
\notag \\
    D_{4} &= \left[0, \dfrac{20}{3}\right]\dfrac{u^4}{12}\eta_1^2\mu^2\Bigg(- \dfrac{1}{\Delta^2} + \dfrac{2}{\Delta}\Bigg),
\notag \\
    D_{5} &= \left[0, \dfrac{20}{3}\right]\dfrac{u^4}{24}\eta_1^2\dfrac{\mu^2}{\alpha}\Bigg(- \dfrac{1}{\Delta^2}
    + \dfrac{2}{\Delta}\Bigg),
\notag \\
    D_6 &= \left[0, \dfrac{20}{3}\right]\dfrac{u^4}{24}\eta_1^2\dfrac{\mu^2}{\alpha}\Bigg(-\dfrac{1}{\Delta^2}
    + \dfrac{1}{\Delta}\left[\dfrac{2}{\alpha} + 3\alpha C_1\right]\Bigg),\notag \\
    D_7 &= \left[16, -4\right]\dfrac{u^4}{48}\eta_1^2\dfrac{\mu^2}{\alpha^2}\Bigg(-\dfrac{1}{\Delta^2}
    + \dfrac{1}{\Delta}\left[\dfrac{2}{\alpha}  + 3\alpha C_1\right]\Bigg).
\end{align}
Here, again, $[a, b]$ represents the factor coming from the traces of $\sigma$-matrices, and $a$ correspond to the contributions  to the
anomalous dimension $\Delta\gamma_{3}$ and $b$ -- to the anomalous dimension $\Delta\widetilde{\gamma}_3$.


\section{Operator-vertex corrections \label{app:ovc}}

Following~\cite{Manashov:2017rrx} we calculate diagrams obtained as a correction to the operator vertex (see Fig.~\ref{fig:OC}), expanding
corresponding $\text{LO}$ contribution, $D_{\text{OC}}(\Delta)$, to the $O(\Delta^0)$ term. For the sum of the diagrams in
Fig.~\ref{fig:OC} one finds
\begin{align}
R^\prime(D)=\frac{D_\text{OC}(\Delta)}\Delta \frac{D_\text{OC}(2\Delta)}{2\Delta}
                            -\frac{D_\text{OC}(0)}\Delta \frac{D_\text{OC}(\Delta)}{\Delta}.
\end{align}

The contribution to the anomalous dimension then reads
For each operator under consideration there are two $\text{LO}$ order diagrams (see Fig.~\ref{fig:LO}),
 so $D_{\text{OC}}(\Delta)/\Delta = D_A(\Delta) + D_B(\Delta)$.
\begin{itemize}
    \item Operator $\pi^2$ and $\pi^{\{a}\pi^{b\}} $
    \begin{align}
        D_A &= -[6, 0]\dfrac{u^2}{6\Delta}\eta_1\mu\Big(1 - 2\Delta(B_2 - 1)\Big),
        \notag \\
        D_B &= -[-2, 4]\dfrac{u^2}{12\Delta}\eta_1\dfrac{\mu}{\alpha}\Bigg\{1 - 2\Delta\Bigg[B_2
        - \dfrac{1}{\alpha} - \dfrac{3}{2}\alpha C_1\Bigg]\Bigg\}.
    \end{align}
    \item Operator $\pi\partial^2\pi$
    \begin{align}
        D_A
        &= -[6]\dfrac{u^2}{12\Delta}\eta_1\left(2\mu - 3\right)\alpha\Bigg\{1
        \notag \\
        &\quad
        -2\Delta\Bigg(B_3 - \dfrac{3}{2} - \dfrac{1}{\mu} - \dfrac{1}{\alpha} + \dfrac{1}{\alpha - 1}\Bigg)\Bigg\},
        \notag \\
        D_B
        &= [-2]\dfrac{u^2}{12\Delta}\frac{\eta_1}\alpha(2\mu - 3)(\mu^2 - 4\mu + 2)\times
        \notag \\
        &\quad
        \Bigg\{1 - 2\Delta\Bigg[B_3 + \dfrac{3\mu\alpha(\alpha - 1)}{2(2\mu - 3)(\mu^2 -4\mu + 2)}C_1
        \notag \\
        &\quad
        - 2 - \dfrac{1}{\mu} - \dfrac{1}{\alpha} + \dfrac{1}{\alpha - 1} - \dfrac{2\mu}{\mu^2 - 4\mu + 2}\Bigg]\Bigg\}.
    \end{align}
\end{itemize}

\begin{figure}[t]
\centering
  \includegraphics[width=0.99\columnwidth]{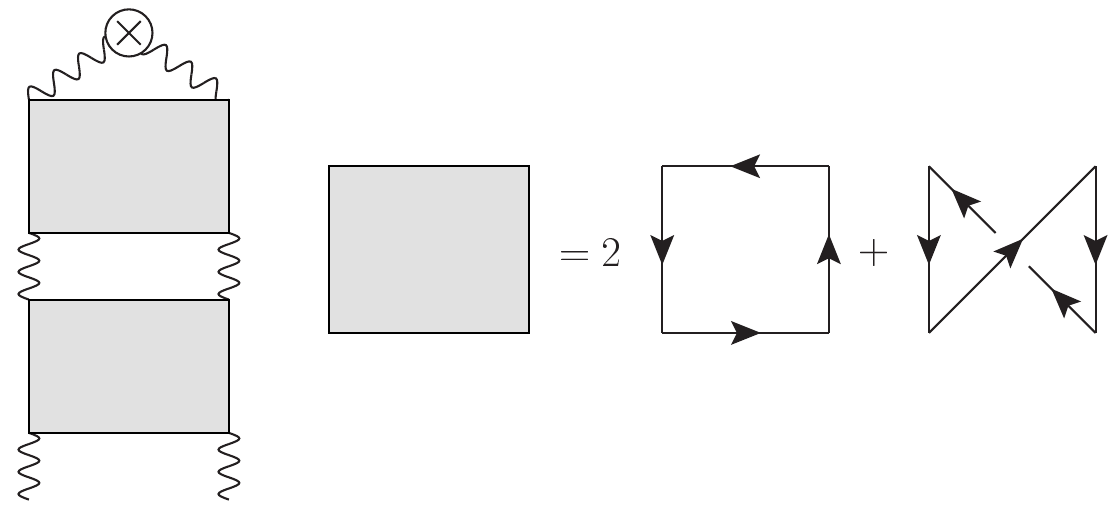}
  \caption{Operator correction diagrams}
  \label{fig:OC}
\end{figure}

\section{Self-energy and vertex correction diagrams \label{app:vertex-se}}

To calculate a contribution from the diagrams obtained as an insertion of $1/n$ SE and vertex corrections into the LO order diagrams
we use the method suggested in~\cite{Derkachov:1997ch}~(see~\cite[Appendix A]{Manashov:2017rrx} for the discussion in Gross-Neveu model).
In this appendix, we present results of relevant calculations.

\subsection{Dressed propagators and vertex}

The expressions for the dressed propagators in $x$-space have the following form ($\kappa = 2\chi$)
\begin{align}
    D_q(x) = -\dfrac{\widehat{A}_q\slashed{x}}{x^{2(\mu + \eta/2)}}, && D^{ab}_{\pi}(x) =
    -\dfrac{\delta^{ab}}{2n}\dfrac{\widehat{B}_{\pi}}{x^{2(1 - \eta - \kappa)}},
\end{align}
where $\kappa = 2\chi$,
\begin{align}
    \widehat{A}_q &= A(\mu)M^{-\eta}\left(1 - \dfrac{\eta}{2\mu}\right),  \\
    \widehat{B}_{\pi} &= B(\mu)M^{2(\eta + \kappa)}\Biggl(1 + \eta \left(B_1 + \frac{1}{\mu}\right)
    \notag
     +\kappa\left(B_1-\frac1\alpha\right)\Biggr).
\end{align}

\begin{figure*}
%
%
\centering
  \includegraphics[width=0.9\textwidth]{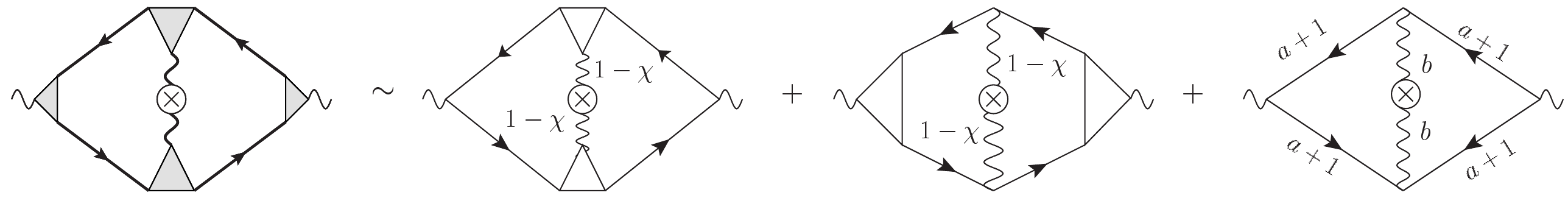}
\caption{An expansion $R(\eta, \chi) = R_1(\chi) + R_2(\chi) + R_3(\eta)$ for the rhombus diagram.
The gray triangles and thick lines on the left-hand side correspond to the dressed vertices and dressed propagators.
 The white triangles on the right-hand side correspond to the dressed vertex~\eqref{dressed-vertex}
 in the limit $\eta \to 0$, and the indices on the lines denote propagator powers.}
 \label{fig:Rhombus}
\end{figure*}
The form of the dressed vertex is fixed by the conformal invariance
\begin{align}
\label{dressed-vertex}
 V(x_1, x_2, x_3) &= \dfrac{\widehat{Z}}{x_{13}^{2(b -\chi)}}
 		      \dfrac{\slashed{x}_{12}}{x_{12}^{2(a + 1 + \chi)}}
 		      \dfrac{\slashed{x}_{23}}{x_{23}^{2(a + 1 + \chi)}}.
\end{align}
where
\begin{align}
    a = \mu - 1 + \eta/2, && b = 1 - \eta,
\end{align}
and
\begin{align}
    \widehat{Z} = \pi^{-2\mu}M^{-2\chi}\,\chi\, 
    a^2(1)\,\alpha^3\left(1 + \dfrac{\eta + \kappa}{\alpha}\right).
\end{align}

To calculate the contribution to the corresponding anomalous dimension, we consider LO diagrams with the insertion of dressed  vertices and
propagators and regulator $\Delta$ in one of the lines. The value of the diagram $D(\Delta)$ then reads
\begin{equation}
    D(\Delta) = \dfrac{R(\eta, \chi)}{\Delta} + O(\Delta^0).
\end{equation}
In $\text{LO}$ there are two diagrams (see Fig.~\ref{fig:LO}), which we refer to as box and rhombus diagrams.

\subsection{Box diagram}
For  the box diagram we obtained
\begin{itemize}
    \item Operators $\pi^2$ and $\pi^{\{a}\pi^{b\}}$ 
    \begin{align}
    R(\eta,\chi) &= -[6, 0]\dfrac{\eta_1}{3}\mu\Big(1 + 2\chi+ \dfrac{3\mu}{2\alpha}\eta\Big),
    \end{align}
    \item Operator $\pi \partial^2 \pi$
\begin{eqnarray}
R(\eta,\chi)\! &=&\! -[6]\dfrac{\eta_1}{3}(2\mu - 3)\alpha\Bigg(1
\notag \\
      \! & \quad&
 + \Bigg[\dfrac{5}{2} + \dfrac{1}{\alpha} - \dfrac{1}{\alpha - 1}
        + \dfrac{2}{2\mu - 3}\Bigg]\chi
\notag\\
\!&\quad&
        +\!\Bigg[\dfrac{7}{8}\! + \dfrac{1}{\alpha}
        + \dfrac{1}{4(\alpha - 1)} +\dfrac{1}{2\mu - 3}\Bigg]\eta\Bigg).
\end{eqnarray}
%
%
\end{itemize}

\subsection{Rhombus diagram}
Calculating
 the rhombus diagram  it is convenient to split it in three part as shown 
 Fig.~\ref{fig:Rhombus}.

\begin{itemize}
    \item Operators $\pi^2$ and $\pi^{\{a}\pi^{b\}} $
\begin{align}
R(\eta,\chi)
&=-[-2, 4]\dfrac{\eta_1}{6}\dfrac{\mu}{\alpha}\Big(1+ {\eta}/{\alpha}
- \left[2/\alpha + 3\alpha C_1\right]{\chi}\Big)
\end{align}
%
%
    \item Operator $\pi \partial^2 \pi$
\begin{align}
R(\eta,\chi)&= [-2]\dfrac{\eta_1}{6}\Bigg\{(\mu^2 -4\mu + 2)(2\mu - 3)/{\alpha}
\notag\\
&\quad
 + \Bigg[4\mu^2 - 6\mu - 11 - \dfrac{2}{\alpha} + \dfrac{1}{\alpha^2} - \dfrac{2}{\alpha - 1}\Bigg]\eta
 \notag\\
 &\quad
  + \Bigg[12\mu^2 - 34\mu - \dfrac{2}{\alpha^2} - \dfrac{4}{\alpha - 1} + 10
  \notag \\& \quad
  + 3\mu(\alpha - 1)C_1\Bigg]\chi\Bigg\},
\end{align}
\end{itemize}

\subsection{Contribution to the anomalous dimensions}
Using the connection between critical indices in chiral Heisenberg model, namely
\begin{equation}
    \chi_1 = - \dfrac{\mu}{6\alpha}\eta_1,
\end{equation}
we derive the following contributions to the anomalous dimensions
\begin{itemize}
    \item Operator $\pi^2$
    \begin{flalign}
        \Delta\gamma^{(2)}_{\pi^2} &=\! -\dfrac{1}{3\alpha}\left(\dfrac{8}{3}
        - 14\mu^2 + \dfrac{10}{3\alpha} + \dfrac{2}{3\alpha^2} + \mu^2C_1\right)\!.
    \end{flalign}
    \item Operator $\pi^{\{a}\pi^{b\}} $
    \begin{align}
        \Delta\gamma^{(2)}_{ab} = \dfrac{4\mu}{9\alpha^3}\left(4\mu - 3 + \dfrac{3}{2}\mu\alpha^2C_1\right).
    \end{align}
    \item Operator $\pi\partial^2\pi$
    \begin{align}
        \Delta\gamma^{(2)}_{-} &= \dfrac{1}{3}\Bigg(15\mu^2 - \dfrac{25}{6}\mu - \dfrac{67}{2} + \dfrac{2}{3\alpha} + \dfrac{8}{3\alpha^2}
         + \dfrac{2}{3\alpha^3}
        \notag \\
        &\quad
        + \dfrac{17}{3(\alpha - 1)} + \left(1 
-\mu\alpha
         + \dfrac{1}{\alpha}\right)C_1\Bigg).
    \end{align}
\end{itemize}


\bibliographystyle{sn-aps}

\bibliography{omega25L}

\end{document}